\documentclass[runningheads]{llncs}
\usepackage{graphicx}
\usepackage{amsmath}
\usepackage{amssymb}

\DeclareSymbolFont{matha}{OML}{txmi}{m}{it}
\DeclareMathSymbol{\varv}{\mathord}{matha}{118}
\usepackage{array}
\usepackage{tabularx}
\usepackage{multirow, booktabs}
\usepackage{float}

\begin{document}
\title{Large Deformation Diffeomorphic Image Registration with Laplacian Pyramid Networks}
\titlerunning{LDDMM with LapIRN}
%
%
%
%

\author{Tony C. W. Mok \and Albert C. S. Chung}
\institute{Lo Kwee-Seong Medical Image Analysis Laboratory,\\
	Department of Computer Science and Engineering,\\
	The Hong Kong University of Science and Technology, Hong Kong }

\maketitle              
\begin{abstract}
Deep learning-based methods have recently demonstrated promising results in deformable image registration for a wide range of medical image analysis tasks. However, existing deep learning-based methods are usually limited to small deformation settings, and desirable properties of the transformation including bijective mapping and topology preservation are often being ignored by these approaches. In this paper, we propose a deep Laplacian Pyramid Image Registration Network, which can solve the image registration optimization problem in a coarse-to-fine fashion within the space of diffeomorphic maps. Extensive quantitative and qualitative evaluations on two MR brain scan datasets show that our method outperforms the existing methods by a significant margin while maintaining desirable diffeomorphic properties and promising registration speed.

\end{abstract}
\section{Introduction}
Deformable registration is the process of computing a non-linear transformation to align a pair of images or image volumes by maximizing certain similarity metric between the images. Deformable registration is crucial in a variety of medical image analysis, including diagnostic tasks, radiotherapy and image-guided surgery. Conventional image registration methods \cite{thirion1998image,ou2011dramms,avants2008symmetric,vercauteren2009diffeomorphic} often rely on the multi-resolution strategy and estimate the target transformation iteratively along with a smoothness regularization. Although conventional image registration methods excel in registration accuracy and diffeomorphic properties (i.e., invertible and topology preserving), the running time of the registration process is dependent on the degree of misalignment between the input images and can be time-consuming with high-resolution 3D image volumes. Recent unsupervised deep learning-based image registration (DLIR) methods \cite{balakrishnan2018unsupervised,dalca2018unsupervised,rohe2017svf,de2017end} have demonstrated promising registration speed and quality in a variety of deformable image registration tasks. They treat the image registration problem as the pixel-wise image translation problem, which attempt to learn the pixel-wise spatial correspondence from a pair of input images by using convolutional neural networks (CNN). This significantly speeds up the registration process and shows immense potential for time-sensitive medical studies such as image-guided surgery and motion tracking. However, these approaches may not be good solutions to unsupervised large deformation image registration for two reasons. First, the gradient of the similarity metric at the finest resolution is rough in general, as many possible transformations of the moving image could yield similar measurements of similarity. Second, the optimization problem without the initialized transformation at the finest resolution is difficult due to the large degrees of freedom in the transformation parameters.


To address this challenge, a preliminary study \cite{de2019deep} proposes to stack multiple CNNs for direct affine and deformable image registrations, which are optimized separately. Zhao et al. \cite{zhao2019recursive} leverage an end-to-end recursive cascaded network to refine the registration result progressively, which is identical to breaking down a large deformation into multiple small deformations. But, both methods are only optimized at the finest level using gradient descent and therefore the results can be sub-optimal as the gradient of the similarity metric can be rough at the finest resolution. Moreover, the recursive cascaded networks consume tremendous extra GPU memory, which limits the possible degree of refinement in 3D settings, resulting in minimal improvement over the brain MR registration tasks. A recent paper \cite{hering2019mlvirnet} avoids these pitfalls and utilizes multiple separated CNNs to mimic the conventional multi-resolution strategy. However, the multiple networks are trained separately and the non-linearity of feature maps in each network are collapsed into a warped image before feeding into the next level. Furthermore, these methods completely ignore the desirable diffeomorphic properties of the transformation, which can further limit their potential for clinical usage.

In this paper, we address the above challenges and present a new deep Laplacian Pyramid Image Registration Network (LapIRN) for large deformation image registration. The main contributions of this work are as follows. We
\begin{itemize}
\setlength\itemsep{0.3em}
\item present a novel LapIRN for large deformable image registration that utilizes the advantages of a multi-resolution strategy while maintaining the non-linearity of the feature maps throughout the coarse-to-fine optimization scheme;
\item propose a new pyramid similarity metric for a pyramid network to capture both large and small misalignments between the input scans, which helps to avoid local minima during the optimization; and
\item present an effective diffeomorphic setting of our method and show that our method guarantees desirable diffeomorphic properties, including the invertibility and topology preservation, of the computed transformations.
\end{itemize}

\section{Methods}
The Laplacian pyramid network has demonstrated its efficiency and effectiveness in a variety of computer vision tasks, including high-resolution image synthetic \cite{wang2018high,denton2015deep}, super-resolution \cite{lai2018fast} and optical flow estimation \cite{xiang2019optical}, in constructing high-resolution solutions, stabilizing the training and avoiding the local minima. Motivated by the successes of Laplacian pyramid networks, we propose LapIRN that naturally integrates the conventional multi-resolution strategy while maintaining the non-linearity of the feature maps throughout different pyramid levels. In the following sections, we describe the methodology of our proposed LapIRN, including the Laplacian pyramid architecture, coarse-to-fine training scheme, the loss function and, finally, we describe the diffeomorphic settings of our method.

\begin{figure}[t]
	\begin{center}
		\includegraphics[width=1.0\linewidth]{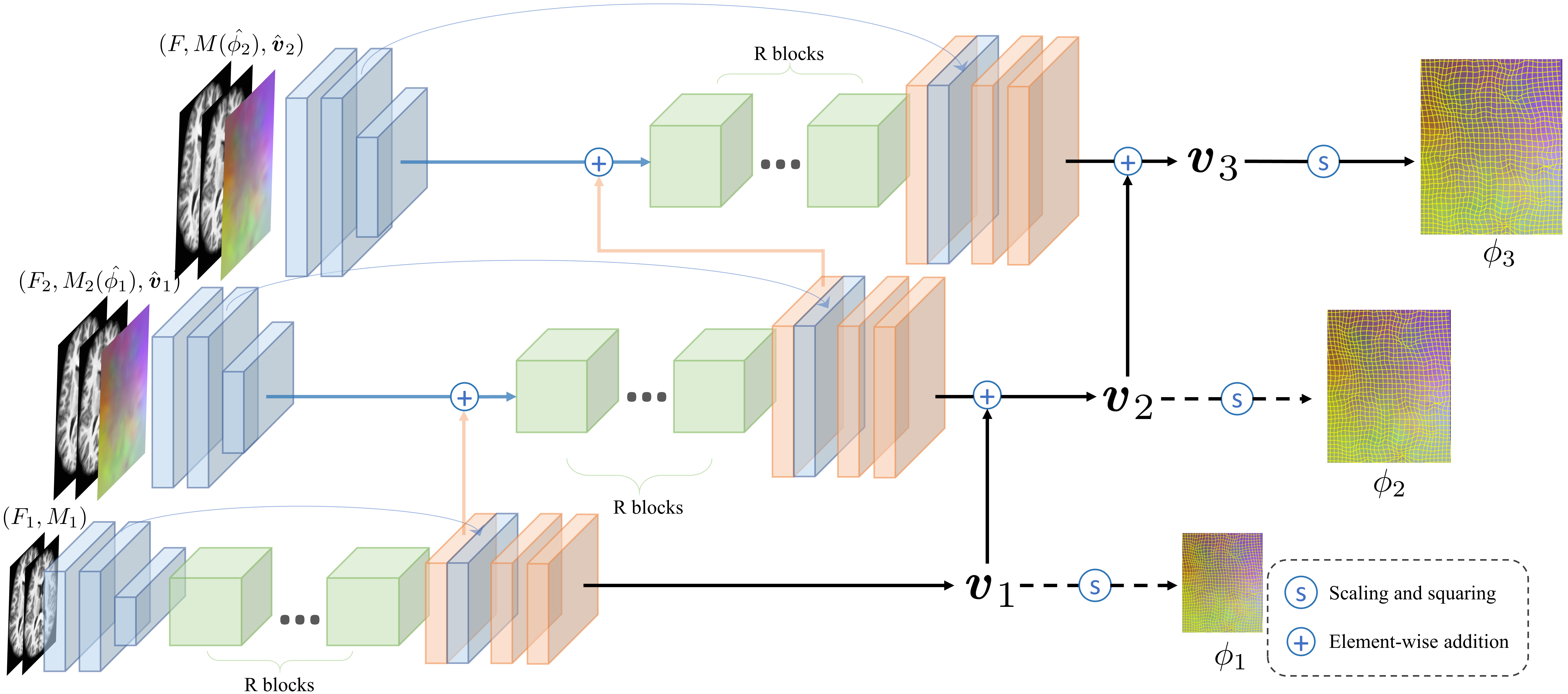}
	\end{center}
	\caption{Overview of the proposed 3-level deep Laplacian pyramid image registration networks in 2D settings. We utilize three identical CNN-based registration networks to mimic the registration with the multi-resolution schema. The feature maps from feature encoder, a set of $R$ residual blocks, and feature decoder are colored with blue, green and red, respectively. The dotted paths are only included in the training phase. We highlight that all registrations are done in 3D throughout this paper. For clarity and simplicity, we depict the 2D formulation of our method in the figure.}
	\label{fig:FCN_archi}
\end{figure}

\subsection{Deep Laplacian Pyramid Image Registration Networks}
Given a fixed 3D scan $F$ and a moving 3D scan $M$, the objective of our method is to estimate a time 1 diffeomorphic deformation field $\phi^{(1)}$ such that the warped moving scan $M(\phi^{(1)})$ is aligned with $F$, subject to the smoothness regularization on the predicted velocity field $\textbf{\textit{v}}$. Specifically, we parametrize the deformable registration problem as a function $f_\theta(F, M) = \phi^{(1)}$ with the Laplacian pyramid framework, where $\theta$ represents the learning parameters in the networks. 

\subsubsection{Network architecture}
We implement our LapIRN using a $L$-level Laplacian pyramid framework to mimic the conventional multi-resolution strategy. For simplicity, we set $L$ to 3 throughout this paper. The overview of LapIRN is illustrated in Fig. 1. Specifically, we first create the input image pyramid by downsampling the input images with trilinear interpolation to obtain $F_i \in \{F_1, F_2,F_3\}$ (and $M_i \in \{M_1, M_2, M_3\}$), where $F_i$ denotes the downsampled $F$ with a scale factor $0.5^{(L-i)}$ and $F_3 = F$. We employ a CNN-based registration network (CRN) to solve the optimization problem for each pyramid level. For the first pyramid level, CRN captures the non-linear misalignment from the concatenated input scans with the coarsest resolution ($F_1$ and $M_1$) and outputs the 3-channel dense vector fields $\textbf{\textit{v}}_1$ and deformation fields $\phi_1$. For pyramid level $i > 1$, we first upsample the output deformation field from the previous pyramid level ($\phi_{i-1}$) by a factor of 2 to obtain $\hat{\phi}_{i-1}$ and warp $M_i$ with $\hat{\phi}_{i-1}$ to obtain a warped moving image $M_i(\hat{\phi}_{i-1})$. Then, we also upsample the output velocity field from the previous level ($\textbf{\textit{v}}_{i-1}$) by a scale factor of 2 (denoted as $\hat{\textbf{\textit{v}}}_{i-1}$) and concatenate it with the input scans ($F_i$ and $M_i(\hat{\phi}_{i-1})$) to form a 5-channel input for the CRN in level $i$. Finally, we add the output velocity fields from level $i$ with upsampled $\hat{\textbf{\textit{v}}}_{i-1}$ to obtain $\textbf{\textit{v}}_i$ and integrate the resulting velocity field to produce the final deformation fields $\phi_i$ for pyramid level $i$. The feature embeddings from CRN at the lower level are added to the next level via a skip connection, which greatly increases the receptive field as well as the non-linearity of the network to learn complex non-linear correspondence at the finer levels.

\subsubsection{CNN-based Registration Network}
The architecture of CRNs is identical among all the pyramid levels. The CRN consists of 3 components: a feature encoder, a set of $R$ residual blocks, and a feature decoder. As shown in Fig. 1,  the feature encoder is comprised of two $3^3$ 3D convolutional layers with stride 1 and one $3^3$ 3D convolutional layer with stride 2. In our implementation, we use 5 residual blocks for each CRN, each containing two $3^3$ 3D convolutional layers with pre-activation structure \cite{he2016identity} and skip connection. Finally, a feature decoder module with one transpose convolutional layer and two consecutive $3^3$ 3D convolutional layers with stride 1, followed by SoftSign activation, is appended at the end to output the target velocity fields $\textbf{\textit{v}}$. A skip connection from the feature encoder to the feature decoder is added to prevent the vanishing of low-level features when learning the target deformation fields. In CRN, each convolution layer has 28 filters and is followed by a leaky rectified
linear unit (LeakyReLU) activation \cite{maas2013rectifier} with a negative slope of 0.2, except for the output convolution layers.

\subsubsection{Coarse-to-fine Training}
Intuitively, our proposed LapIRN can be trained in an end-to-end manner, which is identical to learning a multi-resolution registration with deep supervision \cite{lee2015deeply}. However, we found that end-to-end training for LapIRN is not an ideal training scheme as it is difficult to balance the weight of multiple losses between different resolutions.  To address this issue, we propose to train LapIRN using a coarse-to-fine training scheme with a stable warm start, which is similar to \cite{karras2017progressive,wang2018high}. Specifically, we first train the CRN from the coarsest level alone and then we progressively add the CRN from the next level to learn the image registration problem at a finer resolution. To avoid an unstable warm start, we freeze the learning parameters for all the pre-trained CRNs for a constant $M$ steps whenever a new CRN is added to the training. We set $M$ to 2000 and repeat this training scheme until the finest level is completed.

\begin{figure}[t]
	\begin{center}
		\includegraphics[width=1.0\linewidth]{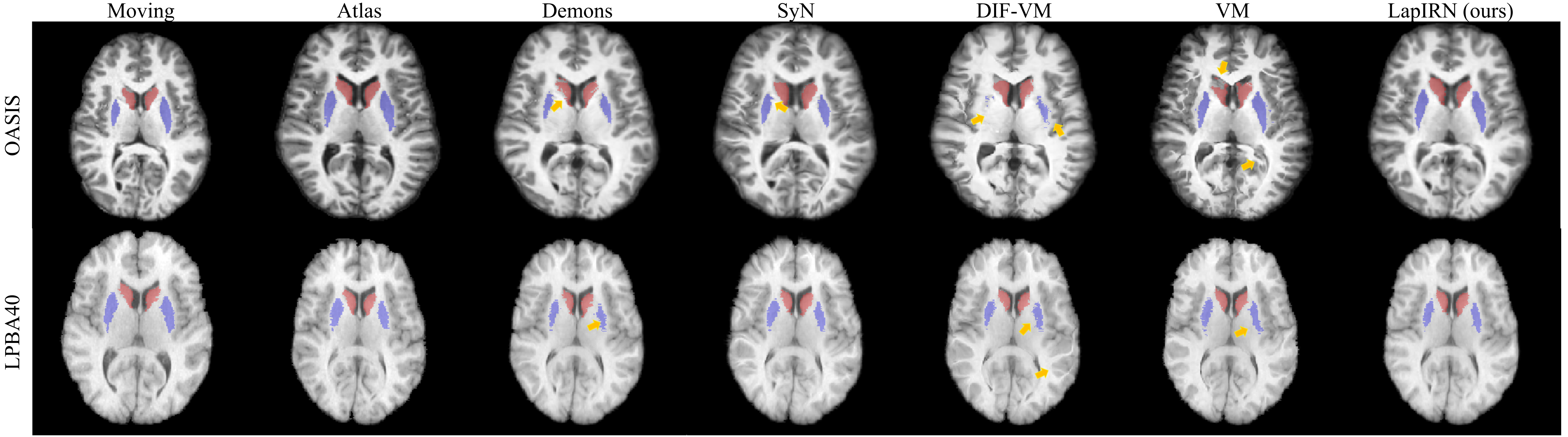}
	\end{center}
	\caption{Example axial MR slices from the moving, atlas and resulting warped images from Demons, SyN, DIF-VM, VM and LapIRN. The caudate and putamen are colored in red and blue respectively. Major artifacts are pointed out by yellow arrows.}
	\label{fig:example_MR}
\end{figure}

\subsection{Similarity Pyramid}
Solving the image registration problem with a intensity-based similarity metric on the finest resolution often results in local minimal solutions. By leveraging the fact that perfectly aligned image pair will yield high similarity values among all resolutions, we propose a similarity pyramid framework to address this challenge. Although the proposed similarity pyramid framework applies to a multitude of similarity measurements, we formulate it using local normalized cross-correlation (NCC) as seen in \cite{balakrishnan2018unsupervised} for simplicity. The proposed similarity pyramid is then formulated as:

\begin{equation}\label{eq:sp}
\mathcal{S}^K(F, M) = \sum_{i \in [1 .. K]} -\frac{1}{2^{(K-i)}} NCC_w(F_i, M_i) ,
\end{equation}

\noindent where $\mathcal{S}^K(\cdot, \cdot)$ denotes the similarity pyramid with $K$ levels, $NCC_w$ represents the local normalized cross-correlation with windows size $w^3$, and $(F_i, M_i)$ denotes the images in the image pyramid (i.e., $F_1$ is the image with the lowest resolution). A lower weight is assigned to the similarity value with lower resolution to prevent the domination of the similarity from lower level. We set $w$ to $1+2i$ in our implementation. The proposed similarity pyramid captures the similarity in a multi-resolution fashion. Since the similarity metric is smoother and less sensitive to noise in the coarser resolution, integrating the similarity metric from a lower level helps to avoid local minima during the optimization problem in high-resolution.

\begin{equation}\label{eq:total_loss}
\mathcal{L}_p(F, M(\phi), \textbf{\textit{v}}) = \mathcal{S}^{p}(F, M(\phi)) + \frac{\lambda}{2^{(L-p)}}||\nabla \textbf{\textit{v}}||^2_2 ,
\end{equation}

\noindent where $p \in [1..L]$ denotes the current pyramid level, the second terms is the smoothness regularization on the velocity field $\textbf{\textit{v}}$, and $\lambda$ is a regularization parameter.

\subsection{Diffeomorphic Deformation}
Recent DLIR methods often parameterize the deformation model using displacement field $u$ such that the dense deformation field $\phi(x) = x + u(x)$, where $x$ represents the identity transformation. Although this parameterization is common and intuitive, the desirable properties of the predicted solution, including topology preservation and invertibility, cannot be guaranteed. Therefore, we parameterize our deformation model using the stationary velocity field under the Log-Euclidean framework and optimize our model within the space of diffeomorphic maps. Specifically, the diffeomorphic deformation field $\phi$ is defined as $\frac{d\phi_t}{dt} = \textbf{\textit{v}} (\phi^{t})$, subject to $\phi^{(0)} = Id$. We follow \cite{arsigny2006log,dalca2018unsupervised} to integrate the (smooth) stationary velocity field $\textbf{\textit{v}}$ over unit time using the scaling and squaring method with time step $T=7$ to obtain the time 1 deformation field $\phi^{(1)}$ such that $\phi^{(1)}$ is approximated to $exp(\textbf{\textit{v}})$, which is a member of the Lie group. Apart from that, we also report the results of LapIRN$_{disp}$, which is a variant of LapIRN parameterizing the deformation model with displacement fields instead.

\section{Experiments}

\subsubsection{Data and Pre-processing}
We have evaluated our method on brain atlas registration tasks using 425 T1-weighted brain MR scans from the OASIS \cite{marcus2007open} dataset and 40 brain MR scans from the LPBA40 \cite{shattuck2008construction} dataset. In the OASIS dataset, it includes subjects aged from 18 to 96 and 100 of the included subjects suffered from very mild to moderate Alzheimer's disease. We carry out standard preprocessing steps, including skull stripping, spatial normalization and subcortical structures segmentation, for each MR scan using FreeSurfer \cite{fischl2012freesurfer}. For OASIS, we utilize the subcortical segmentation maps of the 26 anatomical structures as the ground truth in the evaluation. In the LPBA40 dataset, the MR scans in atlas space and its subcortical segmentation map of 56 structures, which are manually delineated by experts, are used in our experiments. We resample all MR scans with isotropic voxel sizes of $1^3mm$ and center cropped all the preprocessed MRI scans to $144 \times 192 \times 160$. We randomly split the OASIS dataset into 255, 20 and 150 volumes and split the LPBA40 dataset into 28, 2 and 10 volumes for training, validation and test sets, respectively. We randomly select 5 MR scans and 2 MR scans from the test sets as the atlas in OASIS and LPBA40, respectively. Finally, we register each subject to an atlas using different deformable registration methods and list the results in Table \ref{tab:result}. In total, there are 745 and 18 combinations of test scans from OASIS and LPBA40, respectively, included in the evaluation.

\subsubsection{Measurement}
While recent DLIR methods \cite{zhao2019recursive,balakrishnan2018unsupervised,hu2019dual} evaluate their method solely based on the Dice score between the segmentation maps in warped moving scans and the atlas, the quality of the predicted deformation fields, as well as the desirable diffeomorphic properties, are by no means to be ignored. Therefore, we evaluate our method using a sequence of measurements, including the Dice score of the subcortical segmentation maps (DSC), the percentage of voxels with non-positive Jacobian determinant ($|J_\phi|_{\leq0}$), the standard deviation of the Jacobian determinant on the deformation fields (std($|J_\phi|$)), the volume change between the segmentation maps before and after transformation (TC) \cite{rohlfing2003volume}, and the average running time to register each pair of MR scans in seconds (Time), to provide a comprehensive evaluation on registration accuracy and the quality of solutions. 

\begin{figure}[b!]
	\begin{center}
		\includegraphics[width=1.0\linewidth]{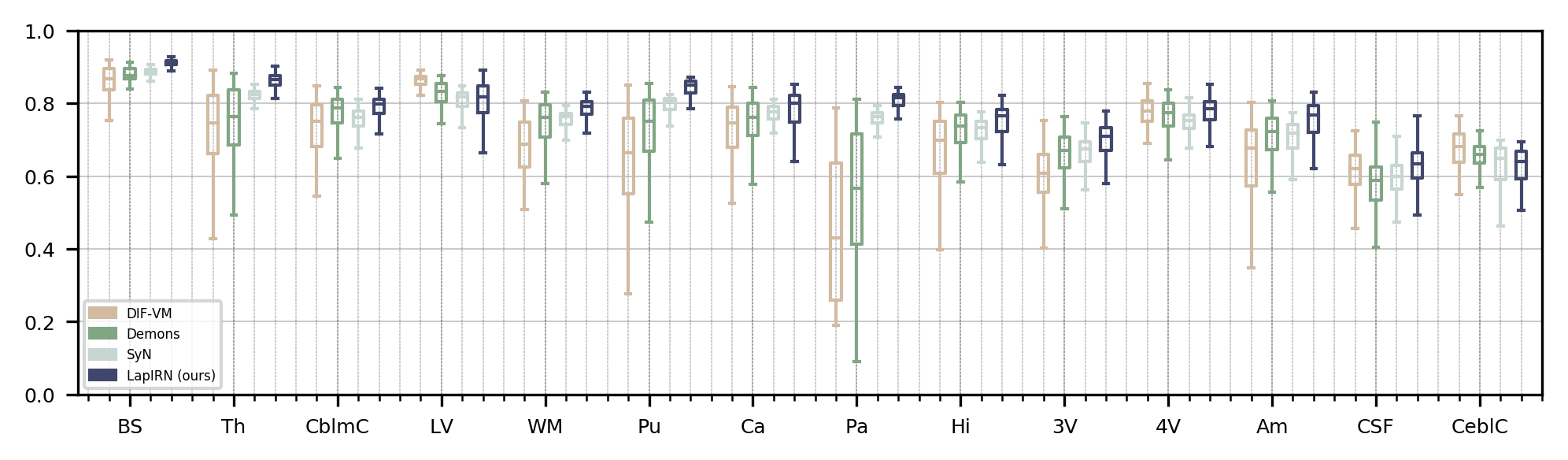}
	\end{center}
	\vspace{-16pt}
	\caption{Boxplots depicting the average Dice scores of each anatomical structure in OASIS for DIF-VM, SyN, Demons and our method. The left and right hemispheres of the brain are combined into one structure for visualization. The brain stem (BS), thalamus (Th), cerebellum cortex (CblmC), lateral ventricle (LV), cerebellum white matter (WM), putamen (Pu), caudate (Ca), pallidum (Pa), hippocampus (Hi), 3rd ventricle (3V), 4th ventricle (4V), amygdala (Am), CSF (CSF), and cerebral cortex (CeblC) are included.}
	\label{fig:box_plot}
\end{figure}

\subsubsection{Implementation}
Our proposed method LapIRN and its variants LapIRN$_{disp}$ are implemented with PyTorch \cite{paszke2017automatic}. We employ an Adam optimizer with a fixed learning rate 1$e^{-4}$. We set $\lambda$ to 4 for LapIRN, which is just enough to guarantee the smoothness of the velocity fields, and $\lambda$ to 1 for LapIRN$_{disp}$. We train our networks from scratch and select the model with the highest Dice score on the validation set.

\subsubsection{Baseline Methods}
We compare our method with two conventional approaches (denoted as Demons \cite{wang2005validation} and SyN \cite{avants2008symmetric}) and two cutting edge DLIR methods (denoted as VM\cite{balakrishnan2018unsupervised} and DIF-VM \cite{dalca2018unsupervised}). Demons and SyN are the top-performing registration among 14 classical non-linear deformation algorithms \cite{klein2009evaluation}. Both Demons and SyN utilize a 3-level multi-resolution strategy to capture large deformation. VM employs a "U" shape CNN structure to learn the dense non-linear correspondence between input scans, while DIF-VM is a probabilistic diffeomorphic variant of VM. For Demons, we use the official implementation in the ITK toolkit \cite{mccormick2014itk}. For SyN, we adopt the official implementation in the ANTs software package \cite{avants2009advanced}. The parameters in Demons and SyN are carefully tuned to balance the tradeoff between registration accuracy and runtime. For the DLIR methods (VM and DIF-VM), we use their official implementation online with default parameters. All DLIR methods are trained from scratch.

\begin{table}[t]
	\centering
	\caption{Quantitative evaluation of the results from OASIS and LPBA40 dataset. DSC indicates registration accuracy. $|J_\phi|_{\leq0}$ represents the average percentage of folding voxels in the deformation fields. std($|J_\phi|$) indicates the smoothness of the deformation fields (lower is better). $TC$ indicates the topology change of the anatomical structure (closer to 1 is better). $Time$ indicates the average running time to register each pair of MR scans in seconds. Initial: spatial normalization.}
	\label{tab:result}
	\makebox[\textwidth][c]{
		\begin{tabular}{ccccccccccc}
			\toprule[1.5pt]
			\multirow{2}{*}{Method} & \multicolumn{5}{c}{OASIS} & \multicolumn{5}{c}{LPBA40} \\
			\cmidrule(lr){2-6}\cmidrule(lr){7-11}
			& \rule{1pt}{0ex} DSC & $|J_\phi|_{\leq0}$ & std($|J_\phi|$) & TC & Time \rule{1pt}{0ex} & \rule{1pt}{0ex} DSC & $|J_\phi|_{\leq0}$ & std($|J_\phi|$) & TC & Time \rule{1pt}{0ex} \\
			\midrule[1pt]
			Initial & 0.567 & - & - & - & - & 0.586 & - & - & - & - \\
			\midrule
			Demons & 0.715 & 0.000 & 0.259 & 1.102 & 192 & 0.720 & 0.048 & 0.174 & 1.004 & 190 \\
			SyN & 0.723 & 0.000 & 0.357 & 1.109 & 1439 & 0.725 & 0.000 & 0.241 & 1.069 & 1225 \\
			DIF-VM & 0.693 & 0.008 & 0.592 & 1.086 & 0.695 & 0.680 & 0.970 & 0.414 & 0.986 & 0.683 \\
			VM & 0.727 & 2.626 & 0.611 & 1.054 & 0.517 & 0.705 & 0.884 & 0.319 & 1.025 & 0.519 \\
			\midrule
			LapIRN$_{disp}$ & 0.808 & 3.031 & 0.651 & 1.161 & 0.312 & 0.756 & 3.110 & 0.728 & 1.033 & 0.310 \\
			LapIRN & 0.765 & 0.007 & 0.319 & 1.101 & 0.331 & 0.736 & 0.008 & 0.301 & 1.032 & 0.334 \\
			\bottomrule[1.5pt]
		\end{tabular}
	}
\end{table}

\subsubsection{Results}
Table \ref{tab:result} gives a comprehensive summary of the results. The variant of our method LapIRN$_{disp}$ achieves 0.808 Dice on a large scale MR brain dataset (OASIS), which outperforms both conventional methods and DLIR methods, Demons, SyN, DIF-VM, VM, by a significant margin of 13\%, 18\%, 17\% and 11\% of Dice score respectively. Nevertheless, similar to methods that work with displacement fields (i.e., VM), the solutions from LapIRN$_{disp}$ and VM cannot guarantee to be smooth and locally invertible as indicated by the high standard deviation of Jacobian determinant (0.65 and 0.61 respectively) and a large percentage of folding voxels (3\% and 2.6\%) in both datasets. Our proposed method LapIRN alleviates this issue and achieves the best registration performance over all the baseline methods, yielding plausible and smooth deformation fields with a standard deviation of the Jacobian determinants of 0.319 and $<0.01 \%$ folding voxels. Furthermore, the inference time of LapIRN is only 0.33 sec, which is significantly faster than the conventional methods (Demons and SyN). We also highlight that our methods outperform the conventional methods even on the small-scale LPBA40 dataset, which has limited training data. Figure \ref{fig:example_MR} illustrates the example of MR slices with large initial misalignment from all methods. The qualitative result shows that LapIRN is capable of large deformation, while the results from VM and DIF-VM are considered to be sub-optimal. Figure \ref{fig:box_plot} depicts the average DSC for each anatomical structure in OASIS dataset. Compare to methods with diffeomorphic properties, our proposed method LapIRN achieves consistently better registration performance among 14 anatomical structures.

\section{Conclusion}
In this paper, we have presented a novel deep Laplacian pyramid networks for deformable image registration with the similarity pyramid, which mimics the conventional multi-resolution strategy to capture large misalignments between input scans. To guarantee the desirable diffeomorphic properties of the deformation fields, we formulate our method with diffeomorphism using the stationary vector fields under the Log-Euclidean framework. Extensive experiments have been carried out and the results showed that not only does our method achieve the state-of-the-art registration accuracy with very efficient running time (0.3 sec), our methods also guarantee desirable diffeomorphic properties of the deformation fields. The formulation of our method can be easily transferred to various applications with minimum effort and has demonstrated immense potentials for time-sensitive medical studies.

\bibliographystyle{splncs04}
\bibliography{myref}

\end{document}